\documentclass[fleqn,twoside]{article}
\usepackage{espcrc2}
\usepackage{pslatex}

\usepackage{graphicx}
\usepackage[figuresright]{rotating}

\newcommand{\AmS}{{\protect\the\textfont2
  A\kern-.1667em\lower.5ex\hbox{M}\kern-.125emS}}

\title{Highly Improved Naive and Staggered Fermions}

\author{Howard D. Trottier\address[CU]{Physics Department, Cornell University,
		Ithaca, NY, 14853, USA}
\address[sfu]{Physics Department, Simon Fraser University,
		Burnaby, B. C., Canada, V5A 1S6, permanent address.},
	G. Peter Lepage\addressmark[CU],
	Paul B. Mackenzie\address{Fermilab, P. O. Box 500, 
		Batavia, IL 60510, USA},
	Quentin Mason\addressmark[CU], and
	Matthew A. Nobes\addressmark[sfu] (HPQCD Collaboration)
}
       
\begin{document}

\begin{abstract}
 We present a new action for highly improved staggered fermions.
 We show that perturbative calculations for the new
 action are well-behaved where those of the conventional staggered action
 are badly behaved.  We discuss the effects of the new terms 
 in controlling flavor mixing, and discuss the design of 
 operators for the action.
\vspace{1pc}
\end{abstract}

\maketitle

\section{INTRODUCTION}

Naive and staggered fermion actions
 (which have very similar properties) have numerous
advantages over actions based on four-component Wilson fermions.  
They have an exact chiral symmetry, allowing a simpler approach to the 
light-quark limit.  They require only even dimension correction operators, 
simplifying the improvement program.  Staggered fermion propagators can be 
calculated much more rapidly than Wilson fermion propagators.

They thus offer the possibility of performing much higher accuracy calculations
on current computers than with other methods.
However, they have historically suffered some drawbacks
which must be addressed before they can be used effectively.
Gluons of momentum $\pi$ induce unphysical transitions between different
flavors of doubler quarks, leading to large flavor symmetry breaking,
especially in the pion sector.  
The perturbation series, even using improved perturbation theory, 
is often more poorly behaved than the series for Wilson fermions.
With staggered fermions, quarks of different spins live on different 
sites of the lattice, leading to large ${\cal O}(a^2)$ errors that are
not present for Wilson fermions.  
Finally, the doubling of the fermion spectrum requires the taking of a root
of the determinant in simulations, 
requiring more care in understanding the locality of the theory.

This paper will show that the improvement program provides an elegant 
solution to
the first three of these problems.

\section{IMPROVED STAGGERED FERMIONS}
One significant problem with ordinary naive and staggered fermions is the large
flavor nondegeneracy.  This problem is worst in the pion sector. 
For the $B$ meson, and other particles with a single light quark, there is no
flavor mixing at the valence level, because the heavy quark would be driven
far off-shell by a gluon of momentum $\pi$.  However, since $B$s couple to
virtual pions, flavor mixing interactions can still be important.

At tree level, they arise from transitions between
doubler quarks of different flavors induced by gluons of momentum 
$\pi$ \cite{lep98,leg98}.
In Ref. \cite{lep99}, it was shown how to turn  ``fat-link'' 
improvement\cite{org98},
which suppresses the coupling of quarks to these gluons,
into a tree level ${\cal O}(a^2)$
improved action by proper tuning of the coefficients and the inclusion of 
an additional term.

%In showing this, we note that
%there is a simple transformation that turns the usual naive fermions $\psi$
%into four copies of staggered fermions $\chi$:
%\begin{equation}
%   \psi(x) \rightarrow \Omega(x) \chi(x), \quad
%   \bar\psi(x) \rightarrow \bar\chi(x) \Omega^\dagger(x) ,
%\end{equation}
%where
%\begin{equation}
%   \Omega(x) = \prod_{\mu=1}^4 (\gamma_\mu)^{x_\mu}.
%\end{equation}

The improved action for naive
quarks is given by:
\begin{equation}
   S_{\rm imp} = \sum_x \bar\psi(x) \left(
   \gamma \cdot \Delta' - {a^2 \over 6} \gamma \cdot \Delta^3 + m
   \right) \psi(x) .
\label{eq:Simp}
\end{equation}
%(There is a simple transformation rewriting the naive action as
%four sets of staggered quarks, so
%most of what we say will be true of both staggered and naive fermions.
%Naive fermions are  simpler to consider.)
(The connection between naive and staggered quarks is the usual 
one \cite{lep99}).
%The improved derivative operator $\Delta'_\mu$ is given by
%\begin{eqnarray}
%   \Delta'_\mu \psi(x) = {1\over 2 a u_0} \bigl(
%   V'_\mu(x) \psi(x+a\hat\mu) \quad \ \nonumber \\
% - V'_\mu(x-a\hat\mu) \psi(x-a\hat\mu) \bigr) ,
%\label{eq:DeltaPrime}
%\end{eqnarray}
%where the $V'_\mu$ link is computed from the true link field
%$U_\mu$ according to
%\begin{equation}
%   V'_\mu(x) = V_\mu(x) - \sum_{\rho\neq\mu}
%   {a^2 (\Delta_\rho)^2 \over 4} U_\mu(x) ,
%\label{eq:VPrime}
%\end{equation}
%with
%\begin{equation}
%   V_\mu = \prod_{\rho\neq\mu}
%   \left( 1 + {a^2 \Delta^{(2)}_\rho \over 4} \right)
 %  \Biggr|_{\rm symm.} U_\mu .
%\label{eq:Vmu}
%\end{equation}
$V_\mu$ is a smeared link variable,
 identical to the ordinary link variable (up to errors of
${\cal O}(a^2)$) for low gluon momentum, but vanishes when a gluon of momentum
$\pi/a$ is extracted.
The action has an exact doubling symmetry\\
%\begin{equation}
$   \psi(x) \rightarrow B_\zeta(x) \psi(x), \quad
   \bar\psi(x) \rightarrow \bar\psi(x) B_\zeta(x) ,
$
%\label{eq:Doubling}
%\end{equation}
where
%\begin{equation}
$   B_\zeta(x) = \prod_\rho
   \left( i \gamma_5 \gamma_\rho \right)^{\zeta_\rho}
   \exp(i \zeta \pi \cdot x / a)
$
%\label{eq:Bzeta}
%\end{equation}
for one of fifteen vectors $\zeta$ with one or more
non-zero components, $\zeta=(1,0,0,0)$, $(1,1,0,0)$, etc.

\section{HIGHLY IMPROVED STAGGERED FERMIONS}

Monte Carlo calculations
 showed a large reduction in pion flavor nondegeneracy with this
action \cite{org99}.  However, significant 
flavor breaking still remains, so further improvement is desirable for
high precision calculations.
In Ref. \cite{dip01}  a nonperturbative determination of gluonic 
corrections to the tadpole improved tree level ${\cal O}(a^2)$ improved staggered
action  was performed.  It found
 a small improvement in pion flavor breaking was possible, but not the large 
reduction that is still desirable.  Therefore, incorporation of additional
operators into the action is needed.

The operators required for full ${\cal O}(\alpha_s a^2)$ improvement 
are four fermi contact interactions \cite{hpl01}.
They arise from transitions in quark--quark scattering between quarks of 
different doubler flavor induced by two gluon exchange.
These contact terms can be transformed into interactions 
between fermion-bilinears and auxiliary scalar fields.
The possible operators are governed by chiral and doubling 
symmetries.
%We consider contact terms of the form $J^2(x)$,
%where $J(x)$ is Hermitian. For each contact interaction one
%introduces a real-valued auxiliary field $\phi(x)$ which has the
%Lagrangian 
%\begin{equation}
%   {\cal L}_\phi = b \phi(x) J(x) + {1 \over 2} \phi^2(x) ,
%\label{eq:Lphi}
%\end{equation}
%without a kinetic term. This is equivalent to the contact term
%\begin{equation}
%   {\cal L}_{\rm contact} = - {1 \over 2} b^2 J^2(x) ,
%\end{equation}
%as can be established by doing the Gaussian integral over $\phi$, or
%using the equations of motion for $\phi$. In a simulation one
%generates an independent Gaussian random distribution for $\phi(x)$
%at each site $x$. The quark is propagated in this background field
%by adding the interaction term in Eq.~(\ref{eq:Lphi}) to the
%tree-level fermion matrix.

%We can therefore remove the one-loop flavor-changing interactions
%generated by the improved quark action Eq.~(\ref{eq:Simp}) using the
%following 

The required 
Lagrangian for the quark-scalar field interactions is
\begin{eqnarray}
   {\cal L}^{\rm (2f)}_\phi(x) \, = \,
   J_\mu^b(x) \Phi_\mu^b(x)  \, +  \, J_\mu(x) \Phi_\mu(x) \quad
   \nonumber \\
  \, + \,  J_{5\mu}^b(x) \Phi_{5\mu}^b(x)  \, +  \, J_{5\mu}(x) \Phi_{5\mu}(x) ,
\label{eq:L2f}
\end{eqnarray}
where $b=1,\ldots,8$ is a color index, and where the $\Phi$ fields are
expressed in terms of normalized auxiliary scalar fields $\phi$
according to
\begin{eqnarray}
   \Phi_\mu^b(x) & = &
   \sum_\zeta c_V^{(8)}(\zeta_\mu) {\cal P}(\zeta) \phi_{\zeta,\mu}^b(x) ,
%   \nonumber \\
%   \Phi_\mu(x) & = &
%   \sum_\zeta c_V^{(1)}(\zeta_\mu) {\cal P}(\zeta) \phi_{\zeta,\mu}(x) ,
%\label{eq:BigPhiV}
\end{eqnarray} etc.
%and
%\begin{eqnarray}
%   \Phi_{5\mu}^b(x) & = &
%   \sum_\zeta c_A^{(8)}(\zeta_\mu) {\cal P}(\zeta) \phi_{\zeta,5\mu}^b(x) ,
%   \nonumber \\
%   \Phi_{5\mu}(x) & = &
%   \sum_\zeta c_A^{(1)}(\zeta_\mu) {\cal P}(\zeta) \phi_{\zeta,5\mu}(x) .
%\label{eq:BigPhiA}
%\end{eqnarray}
The projection operator ${\cal P}(\zeta)$ is given by
\begin{equation}
   {\cal P}(\zeta) = \prod_{\mu=1}^4
   \left( {a^2 \Delta^{(2)}_\mu \over 4} \right)^{\zeta_\mu} .
\label{eq:Pzeta}
\end{equation}
%In the case of the color-octet scalar fields $\phi^b$ the derivatives
%in ${\cal P}(\zeta)$ act according to
%\begin{eqnarray}
%   \Delta^{(2)}_\mu \tilde\phi(x) =
%   {1\over a^2 u_0} \Bigl[
%   U_\mu(x) \tilde\phi(x+a\hat\mu) U^\dagger_\mu(x)
%   - 2 u_0 \tilde\phi(x)
%   \nonumber \\
% + U^\dagger_\mu(x-a\hat\mu) \tilde\phi(x-a\hat\mu) U_\mu(x-a\hat\mu)
%   \Bigr] ,
%\end{eqnarray}
%where
%\begin{equation}
%   \tilde\phi(x) \equiv \sum_b T^b \phi^b(x) .
%\end{equation}
%In the case of the color-singlet scalar fields the operator
%$\Delta^{(2)}_\mu$ acts like an ordinary second derivative, without
%insertion of link variables.
An independent Gaussian random distribution is generated for each
component of each auxiliary field.
% $\phi$ in Eqs.~(\ref{eq:BigPhiV})
%and (\ref{eq:BigPhiA}), normalized according to
%\begin{equation}
%   \left\langle \phi^b_{\zeta,\mu}(x) \phi^{b'}_{\zeta',\mu'} (x') \right\rangle
%   = \delta_{b,b'} \delta_{\zeta,\zeta'} \delta_{\mu,\mu'} \delta_{x,x'} ,
%\label{eq:phiGaussian}
%\end{equation}
%and similarly for the other scalar fields.

The color-octet vector current $J^b_\mu(x)$ for staggered fields is
given by
\begin{eqnarray}
   J^b_\mu(x) = {1 \over 4} \eta_\mu(x) \sum_{\nu=\pm\mu} \Bigl[
   \bar\chi(x) T^b U_\nu(x) \chi(x+a\hat\nu)
   \nonumber \\
   + \, \bar\chi(x+a\hat\nu) U_\nu^\dagger(x) T^b \chi(x) \Bigr] ,
\label{eq:Jmu}
\end{eqnarray}
where $\eta_\mu(x)$ is the usual staggered phase.
%\begin{equation}
%   \eta_\mu(x) = \left( -1 \right)^{x_1 + x_2 + \ldots + x_{\mu-1}} ,
%\label{eq:etaphase}
%\end{equation}
%and where we use the convention
%\begin{equation}
%   U_{-\mu}(x) \equiv U_\mu^\dagger(x-a\hat\mu) .
%\label{eq:Uneg}
%\end{equation}
%Note that the Gell-Mann matrix $T^b$ acts to the left of
%the link in the first term in Eq.~(\ref{eq:Jmu}) and to
%the right of the link in the second term, as required for
%gauge covariance. The expression for the color-singlet
%vector current $J_\mu(x)$ is identical to Eq.~(\ref{eq:Jmu}),
%leaving out the Gell-Mann matrix.

The color-octet axial-vector current $J^a_{5\mu}$ for
staggered fields is given by
\begin{eqnarray}
   J^b_{5\mu}(x) = {1 \over 16} \tau_\mu(x) \sum_\rho \Bigl[
   \bar\chi(x) T^b {\cal U}(x;x+\rho) \chi(x+\rho)
   \nonumber \\
   + \, \bar\chi(x+\rho) {\cal U}^\dagger(x;x+\rho) T^b \chi(x) \Bigr] ,
\label{eq:J5mu}
\end{eqnarray}
where the sum over $\rho$ is over the eight vectors of length
$\sqrt3 a$ perpendicular to $\hat\mu$
\begin{equation}
   \rho^2 = 3a^2 , \quad \rho \cdot \hat \mu = 0 ,
\label{eq:rho}
\end{equation}
and
\begin{eqnarray}
   {\cal U}(x;x+\rho) =
   \Bigl[ U_{\rho_1}(x) U_{\rho_2}(x+a\hat\rho_1) \qquad \qquad & &
   \nonumber \\
   \times
   U_{\rho_3}(x+a\hat\rho_1+a\hat\rho_2) \Bigr]_{\rm symm.} , & &
\label{eq:Urho}
\end{eqnarray}
where the symmetrization is over the six assignments of the components
of $\rho$ to the indices $\rho_1$, $\rho_2$, and $\rho_3$. The
symmetrization is done so as to avoid introducing new discretization
errors. 
The phase $\tau_\mu(x)$ comes from diagonalizing
the axial vector current,
$\bar\psi(x) \gamma_5\gamma_\mu \psi(x+\rho)$.
%$\gamma_5 \gamma_\mu$
%\begin{equation}
%   \tau_\mu(x) = \prod_{\nu\neq\mu} \eta_\nu(x) .
%\label{eq:tauphase}
%\end{equation}
%The expressions for the color-singlet vector and axial-vector currents
%are identical to Eqs.~(\ref{eq:Jmu}) and (\ref{eq:J5mu}), leaving out
%the Gell-Mann matrices.

Finally each of the coefficient functions, 
%e.g. $c^{(1)}_V(\zeta_\mu)$,
%in Eqs.~(\ref{eq:BigPhiV}) and (\ref{eq:BigPhiA}) 
is expressed in
terms of two scalar functions of $\zeta^2$
\begin{eqnarray}
   c^{(8)}_V (\zeta_\mu) & = & \zeta_\mu c^{(8)}_V(\zeta^2)
   + (1 - \zeta_\mu) \tilde c^{(8)}_V(\zeta^2) ,
%   \nonumber \\
%   c^{(1)}_V (\zeta_\mu) & = & \zeta_\mu c^{(1)}_V(\zeta^2)
%   + (1 - \zeta_\mu) \tilde c^{(1)}_V(\zeta^2) ,
\label{eq:cV}
\end{eqnarray}
and analogously for $c^{(1)}_V$,  $c^{(8)}_A$, and $c^{(1)}_A$.
%\begin{eqnarray}
%   c^{(8)}_A (\zeta_\mu) & = & \zeta_\mu c^{(8)}_A(\zeta^2)
%   + (1 - \zeta_\mu) \tilde c^{(8)}_A(\zeta^2)
%   \nonumber \\
%   c^{(1)}_A (\zeta_\mu) & = & \zeta_\mu c^{(1)}_A(\zeta^2)
%   + (1 - \zeta_\mu) \tilde c^{(1)}_A(\zeta^2) .
%\label{eq:cA}
%\end{eqnarray}
The results are given in Table \ref{tab:cImproved}.

\begin{table}[t]
\caption{One-loop coefficients for the fermion bilinears in
Eq.~(\ref{eq:L2f}), using the Symanzik-improved gluon action to compute
the gluon propagator. The results are in units of $\alpha_V(q^*\approx\pi/a)$,
and are accurate to a few parts in the last digit. Entries with an
$i$ are imaginary.}
\label{tab:cImproved}
%\begin{ruledtabular}
\begin{tabular}{l|l|l|l|l}%{c@{\extracolsep{0ptplus1fil}}d@{\extracolsep{0ptplus1fil}}
                %d@{\extracolsep{0ptplus1fil}}d@{\extracolsep{0ptplus1fil}}d}
 $\zeta^2$
 & \multicolumn{1}{r}{$c^{(8)}_V$}  & \multicolumn{1}{r}{$\tilde c^{(8)}_V$}
 & \multicolumn{1}{r}{$c^{(1)}_V$}  & \multicolumn{1}{r}{$\tilde c^{(1)}_V$} \\
 \hline
%      c8V       c~8V     c1V      c~1V
  1  & 0.880 i  & 0.500  & 0.643 i  & 0         \\
  2  & 0.435 i  & 0.438  & 0.217 i  & 0         \\
  3  & 0.335 i  & 0.409  & 0.244 i  & 0         \\
  4  & 0.300 i  & 0      & 0.220 i  & 0         \\
 \hline\hline
 $\zeta^2$
 & \multicolumn{1}{r}{$c^{(8)}_A$}  & \multicolumn{1}{r}{$\tilde c^{(8)}_A$}
 & \multicolumn{1}{r}{$c^{(1)}_A$}  & \multicolumn{1}{r}{$\tilde c^{(1)}_A$} \\
 \hline
%     c8A       c~8A      c1A        c~1A
 1  & 0.404 i  & 0.518    & 0.295 i  & 0         \\
 2  & 0.364    & 0.228 i  & 0        & 0.166 i   \\
 3  & 0.198 i  & 0.198    & 0.145 i  & 0         \\
 4  & 0.190    & 0        & 0        & 0         \\
\end{tabular}
%\end{ruledtabular}
\end{table}

\section{OPERATOR DESIGN: NAIVE VS. STAGGERED FERMIONS}	

Because of the transformation that turns naive fermions into four 
identical copies of staggered fermions, it is clear that a simulation
with naive fermions (taking the sixteenth root of the determinant) and one
with staggered fermions (taking the fourth root) would give identical results.
Staggered fermions are therefore clearly preferable in simulations since they
are a factor of four cheaper.
For valence quarks and their operators, on the other hand,
we should examine each for advantages 
and disadvantages.  
For staggered fermions, different spins live on different lattice sites.
Such operators as the ${\cal O}(a^2)$ improved four-quark operator
used in $B_K$ are spread over $4^4$ blocks on the lattice.
It is reasonable to expect large $(ap)^2$ discretization errors from such
operators, as are indeed observed.
For naive quark operators such as 
$(\overline\psi(x)\Gamma\psi(x))^2$, all quarks live on the same site.
Hence, at tree-level they have no finite $a$ errors to all orders
in $a$.

\section{CONVERGENCE OF PERTURBATION THEORY}

\begin{table}[h]
\caption{For improved staggered quarks (here with Wilson glue), 
unlike unimproved staggered quarks,
all coefficients in one-loop operator renormalizations are of order one.
\label{ops}}
\begin{tabular}{|c|c|c|}
\hline
 & {Unimproved} & Improved  \\
Operator & Staggered &   Staggered \\
\hline
$\bar\psi \psi$ and $\bar\psi \gamma_5 \psi$
  & $1 - 4.17 \alpha_V$  & $1 - 0.39 \alpha_V$ \\
$\bar\psi \gamma^\mu \psi$ and $\bar\psi \gamma^\mu \gamma_5 \psi$
  & $1 - 1.57 \alpha_V$  & $1 + 0.48 \alpha_V$ \\
$\bar\psi \sigma^{\mu\nu} \psi$
  & $1 - 0.56 \alpha_V$ & $1 + 0.66 \alpha_V$ \\
$\bar\psi\bar\psi {\cal O}_+ \psi\psi$
  & $1 - 1.76 \alpha_V$ & $1 + 1.26 \alpha_V$ \\
$\bar\psi\bar\psi {\cal O}_- \psi\psi$
  & $1 - 5.57 \alpha_V$ & $1 + 0.40 \alpha_V$ \\
$\bar\psi\bar\psi {\cal O}_1 \psi\psi$
  & $1 - 2.40 \alpha_V$ & $1 + 1.12 \alpha_V$ \\
$\bar\psi\bar\psi {\cal O}_2 \psi\psi$
  & $1 - 8.11 \alpha_V$ & $1 - 0.17 \alpha_V$ \\
\hline
\end{tabular}
\end{table}

The one case in which the improved perturbation theory program
of Ref. \cite{lep92} does not work well is staggered fermions. 
This was later explained by the existence of fermionic tadpole diagrams
in naive and staggered fermions:\cite{lep98,gol98} fermion loops from all
doubler quark flavors, created by gluon loops of momentum $\pi$.
Since improved staggered fermions suppress precisely the coupling to
these gluons, we should expect that these effects are absent here and
that perturbation theory is  well behaved.
Table \ref{ops} contains one-loop renormalization coefficients for some 
of the naive fermion operators advocated in the previous section.
It is clear that for the improved case, the coefficients of $\alpha$ are all
of order one, as argued, while in the unimproved case they range as high as
eight.

\section{CONCLUSIONS}
We have shown that the improvement program elegantly removes
many of the traditional difficulties of naive and staggered fermions.
In future work, we will expand on the work presented here, and
apply the new action in Monte Carlo calculations.

\end{document}